# Comparative study of the physical properties for the $A_2TiX_6$ ($A=$ Cs or $NH_4$ and $X=$ Cl or Br) vacancy-ordered double perovskites

M. Talebi and A. Mokhtari[*]

Department of Physics, Faculty of Science, Shahrekord University, P. O. Box 115, Shahrekord, Iran

**Abstract**

The vacancy-ordered double perovskites (VODP) are emerging materials for the renewable energy because of their extraordinary stability. In the present work, we have addressed the structural, electronic and optical properties of the $A_2TiX_6$ ($A=$ Cs or $NH_4$ and $X=$ Cl or Br) organic/inorganic halide VODPs based on first-principles calculations. Our results demonstrate that these compounds have several interesting properties, including high stability, suitable band gap and excellent optical absorption. The band structure calculations for the $Cs_2TiBr_6$ and $(NH_4)_2TiBr_6$ perovskites reveal direct band gaps about 1.77 and 1.59 eV respectively, which is predicted these materials be ideal for application in the solar cells. The investigated materials possess high absorption coefficients in the order of $10^5 cm^{-1}$ in the visible light region. Our research can provide a way to identify stable, bio-friendly and high-efficiency light absorber material for use in optoelectronic and photovoltaic devices.

## 1. Introduction

The performance of the organic/inorganic halide perovskites in various fields has impressed the scientific community due to the variety of compounds and structures. The cost-effective synthesis process and excellent light absorption make them ideal to use in the low-cost and high-performance electro-optical devices such as solar cells, light-emitting devices, lasers, and optical detectors [1-6]. The perovskite compounds are known to be excellent light absorbers in solar cells (SC), and the power conversion efficiency (PCE) of single-junction SCs has significantly changed from 3.8% in 2009 to slightly above 25% in recent years [7]. Also the perovskite-based light-emitting diodes (LED) with external quantum efficiencies exceeding 20% have been reported [8]. High-performance optoelectronic devices are mostly based on organic–inorganic lead halide perovskites, which have high optical absorption and emission efficiency [7]. Although the practical application of halide perovskites requires making high-quality and large-scale samples, important issues such as non-toxicity and high stability of the materials are significant, as well. Currently, finding stable and non-toxic perovskites has attracted the attention of researchers. In the last decades, considerable theoretical efforts have been made to understand the fundamental origins of the unique photovoltaic properties of halide perovskites [3, 9-11]. In the current perspective, we have focused on a derivative of the large family of halide perovskites called VODPs, with the general formula of $A_2BX_6$. This structure is derived from doubling the $ABX_3$ unit cell along all three crystallographic axes and removing half of the B cations at the [$BX_6$] clusters center in a checkerboard pattern [12]. Due to the charge neutrality condition, the B-site cations must be tetravalent. These perovskites can also be obtained by considering vacancy instead of B' cations in the $A_2BB'X_6$ double perovskites. In these compounds, [$BX_6$] octahedra cluster are separated from each other [6].

---

[*] Corresponding author (mokhtari@sku.ac.ir).

Most of the previous articles related to the VODPs have been focused on their structural, electronic and dynamic behaviors, while recently the photovoltaic and optoelectronic applications of these materials have been taken into considerations. Lee and co-workers have been studied the Cs2SnI6 perovskite compound, as an applicable structure in the photovoltaic fields [13]. Much researches have been done with the aim of substituting different metals in the $A_2BX_6$ structure, including Sakai et al. [14] work on the $Cs_2PdBr_6$ perovskite in 2017, which introduced this compound as a new perovskite for use in SCs. Likewise, Jo et al. [15] have successfully synthesized a series of Ti-based halide perovskites (HP), including $A_2TiX_6$ (A = $K^+$, $Rb^+$, $Cs^+$ or $In^+$; X = $I$, $Br$ or $Cl$) and $Cs_2TiBr_{6-x}I_x$, using a melt crystallization method and then have investigated these compounds using theoretical approaches. The results indicated that the Ti-based HPs possess desirable attributes, including suitable band gaps, excellent optical absorption and high stability. In particular, this has been shown experimentally that this group of HP can be used for single-junction and tandem photovoltaic applications. Then, they have proceeded to the fabrication of the $Cs_2TiBr_6$ on $TiO_2/FTO$ substrates via a two-step vapor deposition method. The $Cs_2TiBr_6$ films showed an acceptable band gap value of 1.8 eV, elongated and well-adjusted carrier diffusion lengths greater than 100 nm, appropriate energy levels and environmental stability. The $Cs_2TiBr_6$ thin films also exhibited a PCE about 3.3% [16].

In addition, in experimental and theoretical work, Kong et al [17] demonstrated that the titanium-based halide perovskites $Cs_2Ti(Br_xCl_{1-x})_6$, can be synthesized rapidly in a large scale using an aqueous solution-based process at room temperature. The produced materials had tunable band gaps and optical absorption in the visible range and exhibited excellent thermal stabilities. The results provided the low-cost and solution-processed lead-free $Cs_2TiX_6$ perovskites as promising candidates for next-generation eco-friendly optoelectronic devices.

Despite the fact that instability and inadequate tolerance for environmental (e.g., thermal, moisture) stresses has been one the obstacles to the use of organic cations in halide perovskites, the majority of the halide perovskites used in PSC with high power conversion efficiency contain organic cations (e.g., formamidinium ($FA^+$) or methylammonium ($MA^+$)). Therefore, organic cations have always been considered as a favorable candidate instead of the A-site atom in the structure of halide perovskites [4]. In this context, Jo and co-workers [18] described a family of Te-based lead-free perovskites $A_2TeX_6$ (A = MA, FA or BA; X = Br or I, MA = $CH_3NH_3$; FA = $CH(NH_2)_2$; BA = benzylamine). Their investigated compounds were potential materials for optoelectronic devices and showed a wide absorption range of (871-812 nm). Furthermore, these perovskite compounds were found to be robust under ambient conditions and stable for at least two months without showing any signs of phase change.

In another work, Evans et al. [19] using a study of the evolution of the structure and properties of the $A_2PtI_6$, compounds where A = $NH_4^+$, MA, FA and GUA, displayed the importance of hydrogen bonding for stabilizing some of the structures with large A-cations; and from DFT calculations, have found the compound (NH4+)PtI6 has a slightly sharper absorption edge with lower energy than that seen for the other compounds.

In this work, we have used density functional theory [20, 21] to investigate physical properties of the $A_2TiX_6$ (A= Cs or $NH_4$; X= Cl or Br) VODPs. The B-site cation as titanium atom (Ti), in contrast to lead, is a non-toxic, bio-friendly metal and suitable for maintaining the perovskite structures as well. Furthermore, organic or inorganic cations ($Cs^+$ or $NH_4^+$) for the A-site have been considered. Although the $Cs^+$ cation is the largest stable inorganic atom, the optoelectronic properties can be further tuned by replacing the $Cs^+$ cation on this site with an organic ligand such as $NH_4^+$. We have begun with the structural properties, then the electronic and optical properties of these compounds have been investigated, respectively.

## 2. Computational approach

The obtained computational results in this work are based on density functional theory (DFT). We have performed ab-initio calculations using the Quantum Espresso package [22]. The structural optimization has been done by employing PBE-GGA approximation [23] and the ultrasoft pseudopotentials [24]. To increase the accuracy of calculations, we have calculated the ground state energies and the lattice constants using the Murnaghan equation of state [25]. The wave functions and densities of electrons have expanded on a plane-wave basis set to cutoff energy of 65 and 250 Ryd, respectively. The Monkhorst-Pack scheme [26] is regarded to mesh the first Brillouin zone (1BZ) as 4×4×4 k-points. The energy convergence threshold in the self-consistent-field (SCF) iterations is estimated to be $1.0 \times 10^{-8}$ eV and the positions of atoms are fully relaxed until all force values are less than 0.0001 eV/Å.

The calculation of optical properties is based on the time dependent density functional theory (TDDFT) [27], and have performed using the Turbo_eels code as well [28-33]. Optical and ground state calculations have been performed within similar approximations. The Turbo_eels uses the Liouville-Lanczos (LL) approach and calculates the electron energy loss spectroscopy (EELS). To obtain a converged EEL spectrum, the values of Itermax$_0$ = 2000 and Itermax = 10000, are assumed by the constant extrapolation method. In the EELS calculations, the frequency-dependent complex dielectric function ε (ω) has been calculated to simulate the linear optical properties. The imaginary part of the dielectric function, which illustrates the absorption behavior is obtained as follows [34]:

$$\varepsilon_2(\omega) = \frac{Ve^2}{2\pi\hbar m^2 \omega^2} \int d^3k \sum_{c,v} |\langle u_{ck}|M|u_{vk}\rangle|^2 f(u_{ck})(1 - f(u_{vk}))\delta(E_{ck} - E_{vk} - \hbar\omega) \quad (1)$$

Where *e* and *m* represent the charge and mass of electron. The *V, ℏ* and *M* are the volume of the unit cell, the reduced Planck's constant and the momentum operator, respectively. The $u_{ck}$ and $u_{vk}$ stand for the conduction and valence contributions of the Bloch function [35] respectively. Then the real part of the dielectric function $\varepsilon_1(\omega)$ can obtained using the Kramers-Kronig relation [35]

$$\varepsilon_1(\omega) = 1 + \frac{2}{\pi} P \int_0^\infty \frac{\omega' \varepsilon_2(\omega')}{\omega'^2 - \omega^2} d\omega' \quad (2)$$

Where *P* denotes the principal value of an integral. Finally, optical constants such as absorption coefficient α (ω), refractive index n (ω), reflective index R (ω) and energy loss spectrum L (ω) have been derived by using real and imaginary parts of dielectric function as follows [35, 36] :

$$\alpha(\omega) = \sqrt{2}\omega \left[\sqrt{\varepsilon_1^2(\omega) + \varepsilon_2^2(\omega)} - \varepsilon_1(\omega)\right]^{\frac{1}{2}} \quad (3)$$

$$n(\omega) = \left[\sqrt{\varepsilon_1^2(\omega) + \varepsilon_2^2(\omega)} + \varepsilon_1(\omega)\right]^{\frac{1}{2}} / \sqrt{2} \quad (4)$$

$$R(\omega) = \left|\frac{\sqrt{\varepsilon_1(\omega) + i\varepsilon_2(\omega)} - 1}{\sqrt{\varepsilon_1(\omega) + i\varepsilon_2(\omega)} + 1}\right|^2 = \frac{(n-1)^2 + \kappa^2}{(n+1)^2 + \kappa^2} \quad (5)$$

$$L(\omega) = \frac{\varepsilon_2(\omega)}{\varepsilon_1^2(\omega) + \varepsilon_2^2(\omega)} = -\varepsilon_2^{-1}(\omega) \quad (6)$$

### 3. Results and discussions
#### 3.1 Structural properties

In the present work, we have studied the physical properties of the *A$_2$TiX$_6$ (A= Cs or NH$_4$; X= Cl or Br)* VODPs. In these compounds, the geometrical structure contains octahedrons of *Ti (Cl/Br)$_6$* with interstitial sites filled by *Cs$^+$* or *NH$_4^+$* cations. We have simulated these compounds using the DFT approach [20-21] which have been implemented in the Quantum Espresso (QE) package [22]. The overview of conventional structures for these materials are presented in figure 1. The Ti atoms are surrounded by 6 halide ions and also *Cs$^+$* (or *NH$_4^+$*) ions are surrounded by 12 halide ions. These compounds have been usually crystallized in the cubic phase (225_Fm-3m) at the room temperature. The energy-volume optimization curves are obtained using the Murnaghan's equation of state (EOS) [25] and shown in figure 2. It is necessary to mention that, the zero energy has been separately adjusted to the minimum energy of any curve. The calculated equilibrium lattice constant and bulk modulus for all compounds are presented in table 1 and compared with available experimental or theoretical data.

To ensure the structural stability of the investigated perovskite compounds, the cohesive energy, the Goldschmidt tolerance factor ($\tau_F$) and also the octahedral factor have been calculated. The cohesive energy is defined as the total energy of compound minus the sum of the energies of the isolated atoms [37, 38]. In order to calculate an accurate value for the cohesive energy, the energy calculations for isolated atoms and crystal, must be performed at the same level of accuracy. To achieve this requirement, the energy of each atoms have been calculated by considering a large enough cell (FCC structure) containing just one atom. The cell size has been increased so much that the energy convergence with respect to the size of the cell is obtained less than 0.001Ryd. Finally, by dividing the obtained optimized cohesive energy by the total number of atoms of each compound, the average of studied parameter per atom has been calculated and reported (table 1).

The standard value of the tolerance factor that express the structural stability of the cubic phases, is unity and deviated values indicate the structural instability [39]. For the $A_2TiX_6$, the tolerance factor is mathematically formulated as $\tau_F = (r_A + r_x)/\sqrt{2}(r_{Ti} + r_X)$ [40] and also the octahedral factor $\mu$ is calculated as the ratio of $r_{Ti}$ versus $r_X$, where $r_X$, $r_{Ti}$ and $r_A$ are radii of halogens, Ti atom and A-site cations, respectively [39-43]. It is noteworthy to emphasize that according to Shannon ionic radii [44], we have considered constituent ions as packed rigid spheres with radii of 0.65, 1.88, 1.7, 1.81 and 1.96 A° for the $r_{Ti}$, $r_{Cs}$, $r_{NH4}$, $r_{Cl}$ and $r_{Br}$ respectively [44, 45]. Hence, it is noticeable that the tolerance factor is directly dependent on the radius ratios of A-site cation and halide Ti ion. Concerning halide perovskites, the tolerance factor should be subject to the restrictive condition of $0.85 < \tau_F < 1.11$ [41, 42] which implies that the ionic radius of the A cation must be larger than that of the Ti ion. The octahedral factor is often in the range of $0.414 < \mu < 0.592$ [40], however, it has been experimentally demonstrated that a halide perovskite can be synthesized with an octahedral factor lower than 0.414 [42]. The computed values of $\tau_F$ and $\mu$, as presented in table 1, which confirm the structural stability of studied materials.

One of the other approaches to checking the structural stability is the study of phonon frequencies. The behaviors of the phonon dispersion pattern for the $Cs_2TiCl_6$ and $Cs_2TiBr_6$ have been calculated in various literature [46-49] and their dynamical stabilities have been confirmed by obtaining positive phonon frequencies. Therefore considering that our main work has been optical calculations, we have considered dynamical stability as a default by referring to the above works.

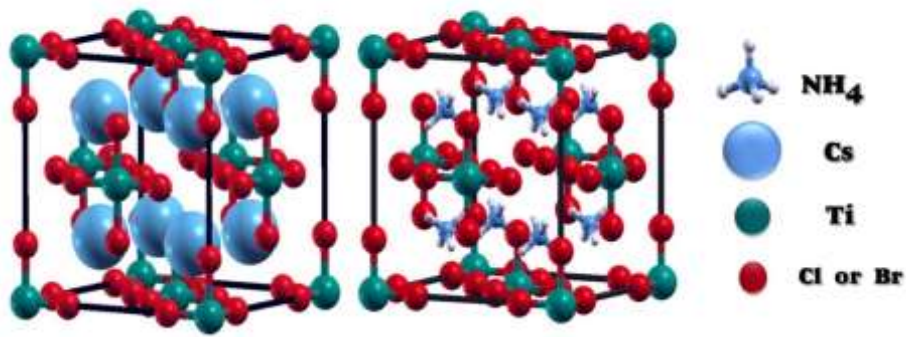

Figure 1: (Color online) The optimized conventional structures for the $Cs_2TiX_6$ (left) and $(NH_4)_2TiX_6$ (right) (X=Cl or Br).

From here on, for simplicity and brevity, the four investigated compounds $Cs_2TiCl_6$, $Cs_2TiBr_6$, $(NH_4)_2TiCl_6$ and $(NH_4)_2TiBr_6$ are named as *CTC*, *CTB*, *NHTC* and *NHTB* respectively. The main features to note from the structural results are as follows:

- The highest percentage of overestimation for the lattice parameters are about 6.10% and 3.52% respect to the available data for *CTC* and *CTB* respectively.
- By comparing the *CTC* and *CTB*, it can be concluded that the smaller lattice constant (therefore smaller average bond length) the larger the bulk modulus (and also larger absolute value of the cohesive energy), consistent with the semi-empirical Cohen's equation [38, 50]. It may also be due to the fact that the electronegativity of Cl and Br atoms are about 3.16 and 2.96 respectively.
- By growing the atomic number from *Cl* to *Br*, the lattice constant as consequently as bond lengths will be increased. This behavior is consistent with Pauling empirical rule that the cation-anion distance is determined by the radius sums [38].
- Our obtained results for the tolerance factors show that all structures are stable. The values of the octahedral factors for the *CTC* and *NHTC* compounds are closer to the determined range [41, 42] and are better than other two compounds. Considering the above results and analyzing the values of cohesive energies, it can be generally said that the CTC compound is a more stable than other compounds and it's results are in good agreement with the others results [17, 46, 51-53].

gure 2: (Color online) The total energies versus primitive unit cell volumes for the $A_2TiX_6$ (A = Cs or $NH_4$, X= Cl or Br) compounds. The solid lines show the Murnaghan EOS fitting. The zero of the energy has been separately set to the minimum energy for each compound.

Table 1: The calculated parameters for the A2TiX6 (A = Cs or $NH_4$, X= Cl or Br) compounds. The available results of others are included for comparison.

| Parameter | $Cs_2TiCl_6$ | $Cs_2TiBr_6$ | $(NH_4)_2TiCl_6$ | $(NH_4)_2TiBr_6$ |
|---|---|---|---|---|
| Lattice constant a ($A^0$) | 10.90, 10.235[b],10.59[c] 10.56[d,e] | 11.08, 10.92[a],11.28[c],10.99[d], 11.025 [e], 10.98[f], 10.69[g] | 10.10 | 10.64 |
| Bulk modulus $B_0$ (GPa) | 25.6, 30.25[c] | 23.2, 23.11[c] | 12.6 | 12.4 |
| Cohesive energy per atom $E_{coh}$(eV/atom) | -3.61 | -3.19 | -3.42 | -3.19 |
| Tolerance factor $\tau_F$ | 1.080 | 1.058 | 1.027 | 1.009 |
| Octahedral factor $\mu$ | 0.334, | 0.309 | 0.334 | 0.309 |
| Band gap (eV) | 2.37, 2.3[b], 2.87[c], 2.219[d] | 1.77, (1.78-2.01)[a] , 1.6[b,h], 1.52[c], 2.0[g], 1.81[g,,k] | 2.17 | 1.59 |
| Static dielectric constant $\varepsilon_1(0)$ | 2.739, 3.07[c] | 3.508, 3.36[c] | 3.283 | 4.072 |
| Static refractive index $n(0)$ | 1.655 | 1.873 | 1.811 | 2.018 |
| Reflectivity $R(0)$ | 0.060 | 0.092 | 0.083 | 0.113 |

Ref[15][a], Ref [17][b], Ref [46][c] ,Ref [51][d], Ref [53][e], Ref [54][f], Ref [55][g], Ref [56,57][h]**,** Ref [58][k]

## 3.2 Electronic properties

In this section, we have investigated the behaviors of the band structures, density of states (DOS), local DOS and charge density of the studied compounds using self-consistent calculations. The band structure behaviors for the all vacancy-ordered compounds have been plotted in the figure 3. As seen in the figure 3, all compounds have direct band gap in the cubic phase at the $\Gamma$ point (The zero energy has been set to the highest occupied state). The valence bands (VB) for all compounds are separated into three sub-bands (VB$_1$, VB$_2$ and VB$_3$: Bottom-up) and their widths are progressively increased. The band gap for all compounds is reported in table 1. The values of 2.37, 1.77, 2.17, and 1.59 eV are estimated for the *CTC*, *CTB*, *NHTC*, and *NHTB* compounds respectively that the first two values are in good agreement with the experimental and theoretical optical band gaps [15, 17, and 46, 51-58]. To the best of our knowledge, no theoretical or experimental results have yet been reported for the values of band gap for the last two compounds. The results show that the band gap increases as the halogen radius decreases. In fact, this gives rise to an increasing in the interaction between halogen and Ti atom which is, in turn, very influential in the band structure properties. In order to study the structures more carefully, their DOS plots have also been attached in figure 3.

The total (whole cell contribution) and partial (site-decomposed local) DOS of all compounds, which are evaluated and analyzed using the tetrahedral method [59], have been plotted in figure 4. To explicitly compare the band-gap difference for the perovskite compounds, the valence band maximum (VBM) for all is adopted as the Fermi level and set to energy zero. It is noticeable that for all compounds, the VBM is primarily occupied by the *X-p (X= Cl or Br)* orbitals while the conduction band minimum (CBM) is mainly dominated by the *Ti-d* orbitals, which is in excellent accordance with previous reports [15, 17, and 46, 51-58]. Meanwhile, the A-site cations (*Cs* or *NH$_4$*) have almost low contributions to the VB and CB.

According to the Shockley-Queisser (SQ) range [60], compounds with an energy gap between 0.9 eV and 1.6 eV can be technologically used in single-junction solar cells. Of course, it is also possible for cases where the band gap is above the upper limit or below the lower limit in the SQ approach to use them in the technologically ideal tandem solar cell [61]. In the present work, with respect to the obtained band gap, we propose the *Cs$_2$TiBr$_6$* and *(NH$_4$)$_2$TiBr$_6$* VODPs, for tandem solar cells.

An overview of charge distribution and bonding among the atoms for the *CTB* and *NHTC* compounds have been presented in figures 5 and 6 in 2-dimentional crystallographic planes (110) and (100). We have not included the results of the other two compounds due to their similarity. It is clear that strong covalent bonding exists between Ti and X (Cl or Br) because charge lines have overlap with each other. The transfer of charge takes place from X to Ti because X is more electronegative than Ti. On other sites, the Cs and X atoms are separate and almost there is not any charge transfer taking place. There are strong covalent bonding exists between N and H in the *NH$_4$* clusters. But in *(NH$_4$)$_2$TiX$_6$* compounds, the valence charges of the *NH$_4$* and X are not separated completely and their chemical bonds can be both ionic and covalent. The ionicity of Cs–Cl bonds is more than the *NH$_4$–Cl* bonds because of the metallic nature of the Cs atoms and also the ionicity of the Cs–Br bonds are also higher than that of the *NH$_4$–Br* bond. The bonding nature can also be explained by the electronegative difference between cations and anions. The calculated electronegativity of Cl, Br, Ti, Cs, H and N are 3.16, 2.96, 1.54, 0.79, 2.20 and 3.04, respectively. If the electronegativity difference between two atoms is less than 1.7, it is a covalent bond and if it is greater than 1.7, it is an ionic bond [62]. Therefore, the electronegativity difference between Ti-Cl (1.62) and Ti-Br (1.42) and N-H (0.84) is smaller than the critical value, which shows strong covalent bonding while Cs-Cl (2.37) and Cs-Br (2.17) are bigger than the critical value that shows the ionic character is dominant.

Figure 3: The calculated band structure and the total DOS for the $A_2TiX_6$ (A = Cs or $NH_4$, X= Cl or Br) using PBE approximation (the top of the valence band is set to zero).

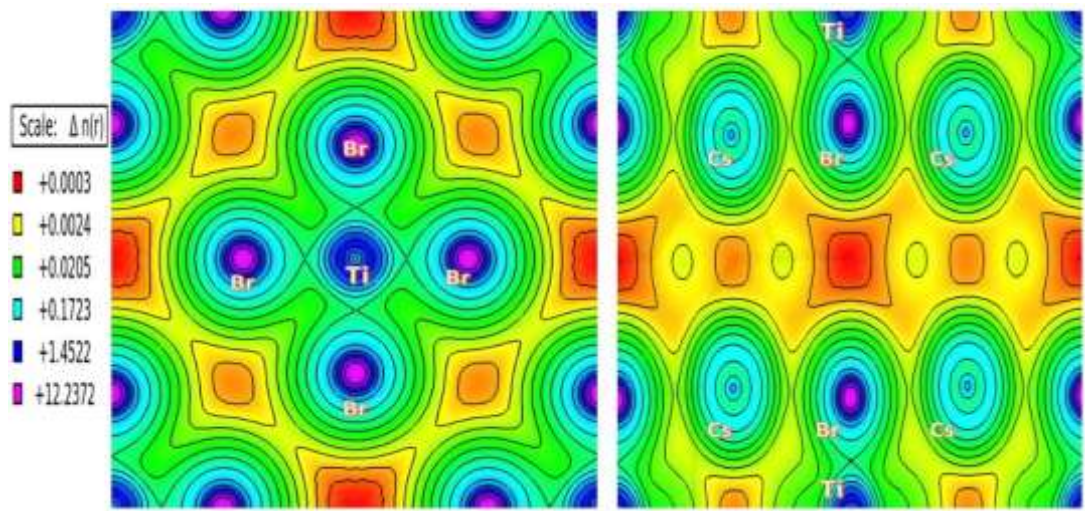

Figure 5: (Color online) The obtained charge density for CTB: the left for plane (100) and the right for plane (110).

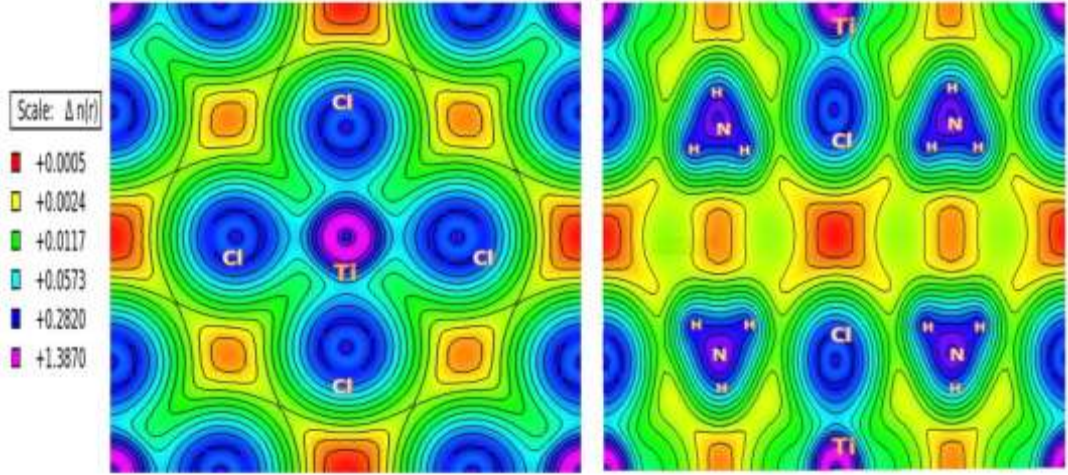

Figure 6: (Color online) The obtained charge density for NHTC in plans (100) and (110), left and right respectively.

### 3.3 optical properties

In the final step, we have investigated the optical properties of the $A_2TiX_6$ ($A=$ Cs or $NH_4$ and $X=$ Cl or Br) VODP using the Liouville-Lanczos (LL) approach [28-33] which is based on the TDDFT [27]. Our calculations have been done using the turbo-eels code that is embedded in the QE package [22]. To probe the optical absorption properties of these materials, it is imperative to introduce the complex dielectric function. In the EELS calculations, the dielectric function $\varepsilon(\omega, \mathbf{q})$ has been calculated to simulate the optical properties. Forasmuch as we have only interested in the low-frequency of optical properties, therefore we have calculated the $\varepsilon(\omega)$ function in the long wavelength limit approximation $\lim_{\mathbf{q} \to 0} \varepsilon(\omega, \mathbf{q})$. We have selected the wave vector in the $X$ [100] direction.

The real and imaginary parts of dielectric function for all compounds are calculated and plotted in figures 7 and 8, respectively. In order to obtain the optimum EEL spectrum, we have optimized two important parameters of the transferred momentum (q) and Lanczos iteration (Itermax$_0$). The spectrums shown in the figures are with optimized parameters of the Itermax$_0$=2000 and q=0.003. According to Eq.2, the real part of the dielectric function $\varepsilon_1(\omega)$ which describes the behavior of light polarization is connected with the imaginary part $\varepsilon_2(\omega)$ which explains the absorption function of the crystal [35].

The real dielectric constant $\varepsilon_1$ increases with enhancement of the photon energy and reaches the maximum value (3.15, 3.99, 3.93, and 4.77) at 2.63 eV, 1.82 eV, 2.38 eV and 1.67 eV for CTC, CTB, NHTC and NHTB, respectively. The maximum peaks reveal that the resonance frequency fully polarized in a perpendicular direction of the electric field of incident light and shift to the lower energy as the cation changed from *Cs* to *NH₄*. Above the resonance peaks, $\varepsilon_1(\omega)$ reduces and energy for all compounds stays positive, which describes the material remains polarized and has semiconducting behavior.

The static dielectric constant $\varepsilon_1(0)$ is associated with band gap ($E_g$) according to Penn's model as follows [63],

$$\varepsilon_1(0) \approx 1 + (\hbar\omega_p/E_g)^2 \qquad (7)$$

Where $\omega_p$ and $\hbar$ are plasma resonance frequency and **reduced** plank's constant, respectively. The static dielectric constant values for all compounds are listed in table 1. As shown, the bromide double perovskites have a larger dielectric constant than the chloride double perovskites, which is also consistent with the obtained band gap values.

As displayed in Fig.8, with respect to the imaginary parts, it is apparent that maximum peaks exist at the range of 0-10 eV and after that, it presents a downward trend. The highest peaks in the $\varepsilon_2(\omega)$ diagrams corresponded with the regions that $\varepsilon_1(\omega)$ from a maximum point starts to decrease to the position that reaches its minimum value. The threshold values for $\varepsilon_2(\omega)$ and the absorption coefficient spectrum are called the base absorption edge. Interband transitions are caused by the excitation of the absorption edge [35]. It can be obtained from Fig.8 that the values of observed thresholds are 2.17 eV, 1.42eV, 1.32eV and 1.92eV for *CTC*, *CTB*, *NHTC* and *NHTB* compounds respectively. Also, maximum values appear at about 6.41, 7.18, 7.87 and 7.81eV respectively. This implies that the investigated compounds have strong absorption ability in the range of the ultraviolet spectrum.

Then, as defined in equations 3-6, using the real and imaginary parts of the dielectric function, we have directly derived other related optical parameters such as absorption coefficient *α (ω)*, refractive index *n (ω)*, reflective index *R (ω)* and energy loss spectrum *L (ω)* (figures 9-12).

The figure 9 displays the calculated absorption coefficients *α(ω)* for *CTC*, *CTB*, *NHTC* and *NHTB* compounds with the absorption edges at 2.27, 1.52, 1.42 and 2.02eV, respectively, which are in reasonable agreement with the corresponding band gaps and is slightly greater than imaginary part of the dielectric function. In the energy windows between 1-15 eV, there are several absorption peaks with increasing trends that can be attributed to the electronic transitions from bonding states to anti-bonding states. In the range of visible energy, the first absorption of perovskites the *CTB* and *NHTB* is at energy 2.69 and 2.32 eV respectively. The first absorption peak of the compound *CTC* and $NH_4TiCl_6$ is at the energy of 3.39 and 3.26 eV, which these energy regions are comprising the visible to near-ultraviolet regions. Therefore, the studied materials are exceedingly significant for solar cells and optoelectronic devices.

The maximum absorption about $127.17 \times 10^4$ cm$^{-1}$ (at 15.79 eV), $122.06 \times 10^4$ cm$^{-1}$ (at 15.51 eV), $136.29 \times 10^4$ cm$^{-1}$ (at 16.77 eV) and also $120.29 \times 10^4$ cm$^{-1}$ (at 16.27 eV) are obtained for CTC, CTB, NHTC and *NHTB*, respectively. After these values, the absorption coefficients decrease to reach a minimum after some fluctuation due to different rates of transition and recombination.

Figure 7: (Color online) The real parts of the dielectric function for the $A_2TiX_6$ (A= *Cs* or *NH$_4$*, X=Cl or Br).

Figure 8: (Color online) The imaginary parts of the dielectric function for the $A_2TiX_6$ (A= Cs or $NH_4$, X=Cl or Br).

Figure 9: (Color online) The absorption coefficient $\alpha(\omega)$ for the $A_2TiX_6$ (A= Cs or $NH_4$, X=Cl or Br).

We have shown the behavior of refractive coefficient as a function of photon energy for the $Cs_2TiX_6$ and $(NH_4)_2TiX_6$ (X= Cl or Br) VODPs in figure 10. It can be observed that the static refractive index is 1.65, 1.81, 1.63 and 2.01 eV for the *CTC*, *CTB*, *NHTC* and *NHTB* perovskites, respectively. The static value of n (0) is related to $\varepsilon_1$ (0) by relation $n^2$ *(0) =* $\varepsilon_1$ *(0)* and hence, computed results are verified (Table 1). The refractive index minimum for the *CTC*, *CTB*, *NHTC* and *NHTB* compounds are occurred in about 24.4, 21.42, 22.96 and 18.79 eV, respectively.

The obtained reflection spectra for the studied compounds have shown in figure 11. With respect to the reflection spectrum, at zero energy, the static reflectivity coefficients have respectively obtained 6.08 %, 9.23%, 8.33% and 11.37 % for the *CTC*, *CTB*, *NHTC* and *NHTB* structures, which are negligible (Table1). There are several peaks in the range of energy 0-17eV and after that, the graphs have a downward trend. The maximum reflectivity values are found for these perovskites about 13.18% (at 15.86 eV), 14.15% (at 7.15 eV), 13.79% (at 16.88 eV) and 16.69% (at 7.90 eV) respectively. These can be concluded that all the studied materials reflect a small percentage of the emitted light.

Finally, the energy loss function *L (ω)* which determines the energy loss of electrons when passing through a material [28-31] have been calculated and plotted in figure 12. The peaks in the loss function are classified into two main types, the plasmon peaks and inter-band transition peaks. The plasmon peaks originate from zeroes of the real part of the dielectric function and the inter-band transition peaks achieve from the most distinguished peaks of the imaginary part of the dielectric function [28].

As shown in Fig.12, there is a maximum peak in the energy loss spectrums at energies about 21.24, 21.27, 22.86 and 19.01 eV for *CTC*, *CTB*, *NHTC* and *NHTB* respectively. Considering that these energies correspond with the minimums given for the refractive index and they are also located in the range where the reflection coefficient diagrams have a downward trend, can be attributed to the plasmonic resonance frequency.

In the visible region, the energy loss coefficient for the studied materials is minimal. In addition, the maximum absorption in the diagrams of imaginary dielectric functions and absorption spectra lies in the energy range of 1-17 eV. In the near of ultraviolet region, the optical energy losses are larger but still lower than in the range of 17 -25 eV where the resonance frequency is located. Therefore the studied materials are more suitable for visible light solar cells, but due to the high absorption coefficient in the range of ultraviolet light, can be used in efficient photovoltaic or other optoelectronic devices that operate in this range.

Figure 10: (Color online) The refractive index $n(\omega)$ for the $A_2TiX_6$ (A= *Cs or NH₄*, X=Cl or Br).

Figure 11: (Color online) The reflective index $R(\omega)$ for the $A_2TiX_6$ (A= *Cs orNH₄* , X=Cl or Br).

Figure 12: (Color online) The energy loss function $L(\omega)$ for the $A_2TiX_6$ (A= *Cs or NH₄*, X=Cl or Br).

**Conclusion**

The structural, electronic and optical properties of the $A_2TiX_6$ ($A=$ Cs or $NH_4$ and $X=$ Cl or Br) organic/inorganic halide VODPs are investigated and analyzed comprehensively by the PBE pseudo-potentials in the framework of DFT. The stability of the cubic phase of these compounds is ensured by calculating the cohesive energy and lying the tolerance factor within the specified range (0.85–1.11). The Ti-X and Cs (or $NH_4$)-X bonds are estimated covalent and ionic, respectively. The studied compounds have a direct band gap between 1.59 and 2.37 eV. Among them, the direct band gap of 1.77 eV and 1.59 eV for *CTB* and *NHTB* make them potential candidates for multi-junction concentrator solar cells. We have also calculated different optical properties in a wide frequency range to obtain useful information about the possible performance of these materials in various optoelectronic devices. The results show that all of the compounds possess optimum electronic and optical properties to be used as visible-light-absorbing materials for photovoltaic applications. The investigated materials possess large absorption coefficients in the order of $10^5$ cm$^{-1}$ in the visible region. The compounds *CTB* and *NHTB*, possess suitable band gaps and relatively high optical absorption as compared to other investigated members of the $A_2BX_6$. Therefore, we hope that the results of our work create motivation for further research to identify sustainable, environmentally friendly and high-efficiency light-absorbing materials as alternatives to the lead-based compounds in the photovoltaic devices, especially solar cells, or other optoelectronic devices.

## Acknowledgments

We acknowledge gratefully the support of the Shahrekord University for this research. Our work is performed in the simulation laboratory of physics department.

## References

[1] N. S. Arul, V. D. Nithya (eds.), Revolution of Perovskite, Materials Horizons: From Nature to Nanomaterials, Springer Nature Singapore Pte Ltd, 2020. https://doi.org/10.1007/978-981-15-1267-41.

[2] S. J. Adjogri, E. L. Meyer, 2020. A Review on Lead-Free Hybrid Halide Perovskites as Light Absorbers for Photovoltaic Applications Based on Their Structural, Optical and Morphological Properties. Molecules. 25(21), e5039. https://doi.org/10.3390/molecules25215039.

[3] X. Wang, T. Zhang, Y. Lou, Y. Zhao, All-inorganic lead-free perovskites for optoelectronic applications, Mater. Chem. Front. 3(3) (2019) 365-375. https://doi.org/10.1039/C8QM00611C .

[4] Q. Chen, N. De Marco, Y. M. Yang, T. B. Song, C. C. Chen, H. Zhao, Z. Hong, Y. Yang, Under the spotlight: The organic–inorganic hybrid halide perovskite for optoelectronic applications. Nano Today. 10(3) (2015) 355-396. https://doi.org/10.1016/j.nantod.2015.04.009.

[5] L. Chu, W. Ahmad, W. Liu, J. Yang, R. Zhang, Y. Sun, X. A. Li, Lead-free halide double perovskite materials: A new superstar toward green and stable optoelectronic applications, Nano-Micro. Lett. 11(1) (2019) 1-18. https://doi.org/10.1007/s40820-019-0244-6.

[6] X. G. Zhao, D. Yang, J. C. Ren, Y. Sun, Z. Xiao, L. Zhang, Rational design of halide double perovskites for optoelectronic applications, Joule. 2(9) (2018) 1662-1673. https://doi.org/10.1016/j.joule.2018.06.017.

[7] NREL, Best research-cell efficiency chart. https://www.nrel.gov/pv/cell-efficiency.html.

[8] K. Lin, J. Xing, L.N. Quan, F.P.G. de Arquer, X. Gong, J. Lu, L. Xie, W. Zhao, D. Zhang, C. Yan, et al. Perovskite Light Emitting Diodes with External Quantum Efficiency Exceeding 20 Per Cent. Nature 562 (2018) 245−248. https://doi.org/10.1038/s41586-018-0575-3

[9] M. Wang, W. Wang, B. Ma, W. Shen, L. Liu, K. Cao, S. Chen, W. Huang, Lead-free perovskite materials for solar cells, Nano-Micro. Lett. 13(1) (2021) 1-36. https://doi.org/10.1007/s40820-020-00578-z

[10] V. Pecunia, L. G. Occhipinti, A. Chakraborty, Y. Pan, Y. Peng, 2020. Lead-free halide perovskite photovoltaics: Challenges, open questions and opportunities. APL Mater. 8(10), 100901. https://doi.org/10.1063/5.0022271.

[11] Y. Wu, X. Li, H. Zen, 2021. Lead-free halide double perovskites: structure, luminescence and applications. Small Struct. 2(3), e2000071. https://doi.org/10.1002/sstr.202000071.

[12] Q. A. Akkerman, L. Manna, What Defines a Halide Perovskite? ACS Energy Lett. 5(2) (2020) 604-610. https://dx.doi.org/10.1021/acsenergylett.0c00039.

[13] B. Lee, C. C. Stoumpos, N. Zhou, F. Hao, C. Malliakas, Yeh, T. J. Marks, M. G. Kanatzidis, and R. P. H. Chang. Air-stable molecular semiconducting iodosalts for solar cell applications: Cs2SnI6 as a hole conductor. J. Am. Chem. Soc. 136 (2014) 15379–15385. https://doi.org/10.1021/ja508464w

[14] N. Sakai, A. A. Haghighirad, M. R. Filip, P. K. Nayak, S. Nayak, A. Ramadan, Z. Wang, F. Giustino, and H. J. Snaith, Solution-processed cesium hexabromopalladate (IV), Cs2PdBr6, for optoelectronic applications. J. Am. Chem. Soc. 139 (2017) 6030–6033. https://doi.org/10.1021/jacs.6b13258

[15] M. G. Ju, M. Chen, Y. Zhou, H. F. Garces, J. Dai, L. Ma, N. P. Padture, X. C. Zeng, Earth-abundant nontoxic titanium (IV)-based vacancy-ordered double perovskite halides with tunable 1.0 to 1.8 eV bandgaps for photovoltaic applications, ACS Energy Lett. 3(2) (2018) 297-304. https://doi.org/10.1021/acsenergylett.7b01167

[16] M. Chen, M. G. Ju, A. D. Carl, Y. Zong, R. L. Grimm, J. Gu, X. C. Zeng, Y. Zhou, N. P. Padture, Cesium titanium (IV) bromide thin films based stable lead-free perovskite solar cells. Joule. 2(3) (2018) 558-570. https://doi.org/10.1016/j.joule.2018.01.009

[17] D. Kong, D. Cheng, X. Wang, K. Zhang, H. Wang, K. Liu, H. Li, X. Sheng, L. Yin, Solution processed lead-free cesium titanium halide perovskites and their structural, thermal and optical characteristics, J. Mater. Chem. C 8.5 (2020) 1591-1597. https://doi.org/10.1039/C9TC05711K

[18] D. Ju, X. Zheng, J. Yin, Z. Qiu, B. Türedi, X. Liu, Y. Dang, B. Cao, o.F. Mohammed, O.M. Bakr, Tellurium-Based Double Perovskites A2TeX6 with Tunable Band Gap and Long Carrier Diffusion Length for Optoelectronic Applications. ACS Energy Lett. 4 (2018) 228–234. https://doi.org/10.1021/acsenergylett.8b02113

[19] H.A. Evans, D.H. Fabini, J.L. Andrews, M. Koerner, M.B. Preefer, G. Wu, R. Seshadri, Hydrogen bonding controls the structural evolution in perovskite-related hybrid platinum (IV) iodides. Inorganic chemistry 57(16) (2018) 10375-10382. https://doi.org/10.1021/acs.inorgchem.8b01597

[20] P. Hohenberg, W. Kohn, 1964. Inhomogeneous electron gas, J. Phys. Rev. 136, e864. https://doi.org/10.1103/PhysRev.136.B864

[21] W. Kohn, L. Sham, 1965. Self-consistent equations including exchange and correlation effects, J. Phys. Rev. 140, e1133. https://doi.org/10.1007/978-3-662-10421-7_30.

[22] P. Giannozzi, et al. Advanced capabilities for materials modelling with Quantum ESPRESSO, J. Phys.: Condens. Matter. 29(46) (2017) 465901. https://doi.org/10.1088/1361-648X/aa8f79

[23] J. P. Perdew, K. Burke, M. Ernzerhof, Generalized gradient approximation made simple, Phys. Rev. Lett. 77 (1996) 3865-3868. http://dx.doi.org/10.1103/PhysRevLett.77.3865

[24] B. Walker, R. Gebauer, 2007. Ultrasoft pseudopotentials in time-dependent density-functional theory, J. Chem. Phys. 127, e164106. http://dx.doi.org/10.1063/1.2786999.

[25] F. D. Murnaghan, The Compressibility of Media under Extreme Pressures Natl. cad. Sci. USA, 30 (1944) 244-247. https://doi.org/10.1073/pnas.30.9.244.

[26] H. J. Monkhorst, J. D. Pack, 1976. Special points for Brillouin-zone integrations, Phys. Rev. B 13 e5188. https://doi.org/10.1103/PhysRevB.13.5188

[27] E. Runge and E. K. U. Gross, 1984. Density-functional theory for time-dependent systems, Phys. Rev. Lett. 52, e997. https://doi.org/10.1103/PhysRevLett.52.997

[28] I. Timrov, 2013. Ab initio study of plasmons and electron-phonon coupling in bismuth: from free-carrier absorption towards a new method for electron energy-loss spectroscopy. PhD Thesis. Cond. Mat. Mtrl-Sci. Ecole Polytechnique X, English. Ffpastel -00823758ff. https://pastel.archives-ouvertes.fr/pastel-00823758

[29] I. Timrov, N. Vast, R. Gebauer, S. Baroni, turboEELS-A code for the simulation of the electron energy loss and inelastic X-ray scattering spectra using the Liouville–Lanczos approach to time-dependent density-functional perturbation theory, Comput. Phys. Comm. 196 (2015) 460-469. http://dx.doi.org/10.1016/j.cpc.2015.05.021

[30] I. Timrov, N. Vast, R. Gebauer, S. Baroni, 2013. Electron energy loss and inelastic x-ray scattering cross sections from time-dependent density-functional perturbation theory, Phys. Rev B, 88(6) e064301. https://doi.org/10.1103/PhysRevB.88.064301

[31] I. Timrov, M. Markov, T. Gorni, M. Raynaud, O. Motornyi, R. Gebauer, S. Baroni, N. Vast, 2017. Ab initio study of electron energy loss spectra of bulk bismuth up to 100 ev, Phys. Rev. B 95(9) e094301. https://doi.org/10.1103/PhysRevB.95.094301.

[32] O. Motornyi, Ab initio study of electronic surfaces states and plasmons of gold: role of the spin-orbit coupling and surface geometry, 2018. PhD Thesis., University Paris-Saclay (ComUE).

[33] O. Motornyi, N. Vast, I. Timrov, O. Baseggio, S. Baroni, A. Dal Corso, 2020. Electron energy loss spectroscopy of bulk gold with ultrasoft pseudopotentials and the Liouville-Lanczos method, Phys. Rev. B, 102.3, 035156. https://doi.org/10.1103/PhysRevB.102.035156.

[34] W. Li, S. Zhu, Y. Zhao, Y. Qiu, 2020. Structure, electronic and optical properties of $Cs_2Ti(Br_{1-x}Y_x)_6$ (Y= Cl, I; x= 0, 0.25, 0.5, 0.75, 1) perovskites: the first principles investigations, J. Solid State Chem. 284, e121213. https://doi.org/10.1016/j.jssc.2020.121213

[35] M. Dresselhaus, G. Dresselhaus, S. B. Cronin, A. G. Souza Filho, Optical Properties of solids In: Solid State Properties - Part 3, Alemania: Springer-Verlag, 2018.

[36] M. Talebi, A. Mokhtari, V. Soleimanian, (2023). Ab-inito simulation of the structural, electronic and optical properties for the vacancy-ordered double perovskites $A_2TiI_6$ (A= Cs or NH4); a time-dependent density functional theory study. J. Phys. Chem. Solids, 111262. https://doi.org/10.1016/j.jpcs.2023.111262

[37] T. Hashemifar, A. Mokhtari, V. Soleimanian, Electronic, Structural and Magnetic Properties of the $Sr_2CoWO_6$ Double Perovskite Using GGA +U, J. Super. Nov. Mag. 30 (2017) 497-503. https://doi.org/10.1007/s10948-016-3793-7

[38] A. Mokhtari, 2008. Density functional study of the group II phosphide semiconductor compounds under hydrostatic pressure, J. Phys. Condens. Matter. 20, e135224. https://doi.org/10.1088/0953-8984/20/13/135224

[39] W. Rahim, A. Cheng, C. Lyu, T. Shi, Z. Wang, D. O. Scanlon, R. G. Palgrave, Geometric Analysis and Formability of the Cubic $A_2BX_6$ Vacancy-Ordered Double Perovskite Structure, Chem. Mater. 32(22) (2020) 9573-9583. https://doi.org/10.1021/acs.chemmater.0c02806.

[40] A. E. Fedorovskiy, N. A. Drigo, M. K. Nazeeruddin, 2020. The role of Goldschmidt's tolerance factor in the formation of $A_2BX_6$ double halide perovskites and its optimal range. Small Methods. 4(5), e1900426. https://doi.org/10.1002/smtd.201900426

[41] Z. Xiao, Y. Yan, 2017. Progress in theoretical study of metal halide perovskite solar cell materials, Adv. Energy Mater. 7.22, e1701136. https://doi.org/10.1002/aenm.201701136.

[42] C. Li, X. Lu, W. Ding, L. Feng, Y. Gao, Z. Guo, Formability of $ABX_3$ (X = F, Cl, Br, I) halide perovskites, Acta Crys. Sect. B: Struct. Sci. 64 (2008) 702-707. https://doi.org/10.1107/S0108768108032734

[43] H. A. Maddah, V. Berry, S. K. Behura, S. K. 2020. Cuboctahedral stability in Titanium halide perovskites via machine learning, Comput. Mater. Sci. 173, 109415. https://doi.org/10.1016/j.commatsci.2019.109415.

[44] R. D. Shannon, Revised effective ionic radii and systematic studies of interatomic distances in halides and chalcogenides, Acta Crystallogr. Sect. A 32(5) (1976) 751-767. https://doi.org/10.1107/S0567739476001551

[45] V. Sidey, On the effective ionic radii for ammonium, Acta Crystallogr., Sect. B: Struct. Sci., Cryst. Eng. Mater.72 (4) (2016) 626-633. http://dx.doi.org/10.1107/S2052520616008064

[46] Q. Mahmood, M. Hassan, N. Yousaf, A.A. AlObaid, T.I. Al-Muhimeed, M. Morsi, H. Albalawi, O.A. Alamri, (2022). Study of lead-free double perovskites halides $Cs_2TiCl_6$, and $Cs_2TiBr_6$ for optoelectronics, and thermoelectric applications. Mater. Sci. Semicond. Process. 137, 106180. https://doi.org/10.1016/j.mssp.2021.106180

[47] B. Cucco, G. Bouder, L. Pedesseau, C. Katan, J. Even, M. Kepenekian, G. Volonakis, 2021. Electronic structure and stability of $Cs_2TiX_6$ and $Cs_2ZrX_6$ (X= Br, I) vacancy ordered double perovskites. Appl. Phys. Lett. 119(18), e181903. https://doi.org/10.1063/5.0070104.

[48] D. Liu, W. Zha, R. Yuan, J. Chen, R. Sa, A first-principles study on the optoelectronic properties of mixed-halide double perovskites $Cs_2TiI_{6-x}Br_x$, New J. Chem. 44(32) (2020) 13613-13618. https://doi.org/10.1039/D0NJ02535F

[49] D. Liu, R. Sa, 2020. Theoretical study of Zr doping on the stability, mechanical, electronic and optical properties of $Cs_2TiX_6$, Opt. Mater. 110, e110497. https://doi.org/10.1016/j.optmat.2020.110497

[50] M. L. Cohen, (1985). Calculation of bulk moduli of diamond and zinc-blende solids. Phys. Rev. B 32 (12) 7988. https://doi.org/10.1103/PhysRevB.32.7988

[51] W. Li, S. Zhu, Y. Zhao, Y. Qiu, 2020. Structure, electronic and optical properties of $Cs_2Ti(Br_{1-x}Y_x)_6$ (Y= Cl, I; x= 0, 0.25, 0.5, 0.75, 1) perovskites: the first principles investigations, J. Solid State Chem. 284, e121213. https://doi.org/10.1016/j.jssc.2020.121213

[52] G.K. Grandhi, A. Matuhina, M. Liu, S. Annurakshita, H. Ali-Löytty, G. Bautista, P. Vivo, (2021). Lead-free cesium titanium bromide double perovskite nanocrystals. Nanomaterials, 11(6), 1458. https://doi.org/10.3390/nano11061458

[53] C. Kaewmeechai, Y. Laosiritaworn and A. P. Jaroenjittichai, (2021). First-principles study on structural stability and reaction with H2O and O2 of vacancy-ordered double perovskite halides: Cs2 (Ti, Zr, Hf) X6 Results Phys. **25** (2021) 104225. https://doi.org/10.1016/j.rinp.2021.104225

[54] M. Tsuyama, S. Suzuki, 2019. First-Principles Study on Electronic and Optical Properties of Pb-Free Halide Perovskites $Cs_2TiX_6$ (X= Br, I) in Comparison with $CH_3NH_3PbX_3$ (X= Br, I), J. Phy. Soc. Jpn. 88(10), e104802. https://doi.org/10.7566/JPSJ.88.104802.

[55] J. Euvrard, X. Wang, T. Li, Y. Yan, D. B. Mitzi, Is $Cs_2TiBr_6$ a promising Pb-free perovskite for solar energy applications? J. Mater. Chem. A 8(7) (2020) 4049-4054. https://doi.org/10.1039/C9TA13870F

[56] K. Chakraborty, S. Paul., U. Mukherjee, S. Das. 2021. Impact of Absorbing Layer Band Gap and Light Illumination on the Device Performance of a Single Halide $Cs_2TiX_6$ Based PSC, J. Nano- Electron. Phys. 13(3), e03009. https://doi.org/10.21272/jnep.13(3).03009.

[57] K. Chakraborty, M. G. Choudhury, S. Paul, Numerical study of Cs2TiX6 (X= Br−, I−, F− and Cl−) based perovskite solar cell using SCAPS-1D device simulation, Sol. Energy 194 (2019) 886-892. https://doi.org/10.1016/j.solener.2019.11.005

[58] D. Liu, W. Zha, R. Yuan, J. Chen, R. Sa, A first-principles study on the optoelectronic properties of mixed-halide double perovskites $Cs_2TiI_{6-x}Br_x$, New J. Chem. 44(32) (2020) 13613-13618. https://doi.org/10.1039/D0NJ02535F

[59] M. Kawamura, Y. Gohda, S. Tsuneyuki, 2014. Improved tetrahedron method for the Brillouin-zone integration applicable to response functions, Phys. Rev. B 89, e094515. https://doi.org/10.1103/PhysRevB.89.094515

[60] W. Shockley, H. J. Queisser, 1961. Detailed Balance Limit of Efficiency of p-n Junction Solar Cells, J. Appl. Phys. 32, e510. https://doi.org/10.1063/1.1736034

[61] T. Kirchartz, U. Rau, 2018. What makes a good solar cell? Adv. Energy Mater. 8(28), e1703385. https://dx.doi.org/10.1021/acsenergylett.0c00039

[62] L. Pauling, (1932). The nature of the chemical bond. IV. The energy of single bonds and the relative electronegativity of atoms J. Am. Chem. Soc. 54(9), 3570-3582. https://doi.org/10.1021/ja01348a011.

[63] D.R. Penn, Wave-number-Dependent dielectric function of semiconductors. Phys. Rev. 128 (1962) 2093. https://doi.org/10.1103/PhysRev.128.2093